\documentclass[a4paper]{jpconf}

\usepackage{amsmath}
\usepackage{amssymb}
\usepackage{amsthm}
\usepackage{graphicx}
\usepackage{url}

\newcommand{\mi}{\mathrm{i}} 

\newcommand{\dfd}[3]{\hspace{-0.4em}\ensuremath{\frac{\mathrm{d}^{#1}#3}{(2\pi)^{#2}}}\,}

\newcommand{\eqn}[1]{Eq.~(\ref{#1})}
\newcommand{\fig}[1]{Fig.~\ref{#1}}
\newcommand{\tab}[1]{Table~\ref{#1}}

\newcommand{\eqns}[1]{Eqs.~(\ref{#1})}

\def\Kbar{\overline{K}}
\def\K0bar{\overline{K^0}}

\begin{document}

\title{
  \hfill{\small LFTC-17-11/11} \\
  \vspace*{25mm}
  Partial restoration of chiral symmetry in cold nuclear matter: the
  $\phi$-meson case}

\author{J J Cobos-Mart\'{\i}nez$^{1,2,}$\footnote[5]{Speaker}, K Tsushima$^{2}$,
G Krein$^{3}$, and A W Thomas$^{4}$}

\address{$^{1}$CONACyT--Departamento de F\'{\i}sica, Centro de
Investigaci\'on y de Estudios Avanzados del Instituto Polit\'ecnico
Nacional, Apartado Postal 14-740, 07000, Ciudad de M\'exico, M\'exico.}
\address{$^{2}$Laborat\'orio de F\'{\i}sica Te\'orica e Computacional-LFTC,
  Universidade Cruzeiro do Sul, 01506-000, S\~ao Paulo, SP, Brazil}
\address{$^{3}$Instituto de F\'{\i}sica Te\'orica, Universidade Estadual
  Paulista, Rua Dr. Bento Teobaldo Ferraz, 271-Bloco II, 01140-070, S\~ao
  Paulo, SP, Brazil}
\address{$^{4}$ARC Centre of Excellence for Particle Physics at the Terascale
  and CSSM, Department of Physics, University of Adelaide, Adelaide SA 5005,
  Australia}

\ead{jcobos@fis.cinvestav.mx}

\begin{abstract}

  The work presented at this workshop is divided into two parts. In the first
  part, the mass and decay width of the $\phi$-meson in cold nuclear matter are
  computed in an effective Lagrangian approach. The medium dependence of these
  properties are obtained by evaluating kaon-antikaon loop contributions to
  the $\phi$-meson self-energy, employing medium-modified kaon masses
  calculated using the quark-meson coupling model. The loop integral is
  regularized with a dipole form factor, and the sensitivity of the results
  to the choice of cutoff mass in the form factor is investigated. At normal
  nuclear matter density, we find a downward shift of the $\phi$ mass by a
  few percent, while the decay width is enhanced by an order of magnitude.
  Our results support the literature which suggest that one should observe a
  small downward mass shift and a large broadening of the decay width. In the
  second part, we present $\phi$-meson--nucleus bound state energies and
  absorption widths for four selected nuclei, calculated by solving the
  Klein-Gordon equation with complex optical potentials. The attractive
  potential for the $\phi$-meson in the nuclear medium originates from the
  in-medium enhanced KK loop in the $\phi$-meson selfenergy. The results
  suggest that the $\phi$-meson should form bound states with all the nuclei
  considered. However, the identification of the signal for these predicted
  bound states will need careful investigation because of their sizable
  absorption widths.
  
\end{abstract}

\section{Introduction}

The properties of light vector mesons at finite baryon density, such as their
masses and decay widths, have attracted considerable experimental and
theoretical interest over the last few decades~\cite{Hayano:2008vn,Leupold:2009kz,Hosaka:2016ypm,Krein:2017usp,Metag:2017yuh}. This has been in part
due to their potential to carry information on the partial restoration of
chiral symmetry, and the possible role of QCD of van der Waals forces in the
binding of quarkonia to nuclei. In particular, there is special interest on
the $\phi$-meson, the main reasons being: i) despite its nearly pure
$s\overline{s}$ content, the $\phi$-meson does interact strongly with a
nucleus, composed predominantly of light $u$ and $d$ quarks, through the
excitation of below-threshold virtual kaon and anti-kaon states that might
have their properties changed in medium~\cite{Tsushima:1997df,Laue:1999yv,SchaffnerBielich:1999cp,Akaishi:2002bg,Fuchs:2005zg}; (ii) the $\phi N$ interaction
in vacuum~\cite{Titov:1997qz,Titov:1998bw,Oh:1999nv,Oh:2001bq}and a possible in-medium mass shift of the $\phi$ are related to the strangeness content of the nucleon~\cite{Gubler:2014pta}, which may have
implications beyond the physics of the strong interaction~\cite{Bottino:2001dj,Ellis:2008hf,Giedt:2009mr};
(iii) medium modifications of $\phi$-meson properties have been proposed~\cite{Sibirtsev:2006yk}
as a possible source for the anomalous nuclear mass number $A$-dependence
observed in $\phi$-meson production from nuclear targets~\cite{Ishikawa:2004id};
(iv) furthermore, as the $\phi$-meson is a nearly pure $s\overline{s}$ state
and gluonic interactions are flavor blind, studying it serves to test theories
of the multiple-gluon exchange interactions, including long range QCD van der
Waals forces~\cite{Appelquist:1978rt}, which are believed to play a role in the binding of the
$J/\Psi$ and other exotic heavy-quarkonia to matter~\cite{Brodsky:1989jd,Luke:1992tm,Sibirtsev:1999jr,Gao:2000az,Beane:2014sda,Brambilla:2015rqa,Gao:2017hya,Kawama:2014pja,Kawama:2014iwa,Ohnishi:2014xla,Aoki:2015qla,Morino:2015bqa}. Heavy-ion
collisions and photon- or proton-induced reactions on nuclear targets have
been used to extract information on the in-medium properties of hadrons.
Several experiments have focused on the light vector $\rho$, $\omega$, and
$\phi$ mesons, since their mean-free paths can be comparable with the size
of a nucleus after being produced inside the nucleus. However, a unified
consensus has not yet been reached among the different experiments--see
Refs.~\cite{Hayano:2008vn,Leupold:2009kz,Hosaka:2016ypm,Krein:2017usp,Metag:2017yuh} for comprehensive reviews on the current status.
For the $\phi$-meson, although the precise values are different, a large
in-medium broadening of its decay width has been reported by most of the
experiments performed, while only a few of them find evidence for a
substantial mass shift~\cite{Ishikawa:2004id,Muto:2005za,Mibe:2007aa,Qian:2009ab,Wood:2010ei,Polyanskiy:2010tj}. In 2007 the KEK-E325 collaboration
reported a 3.4\% mass reduction of the $\phi$-meson~\cite{Muto:2005za} and an in-medium
decay width of $\approx$ 14.5 MeV at normal nuclear matter density
$\rho_{0}= 0.15$ fm$^{-3}$. These conclusions were based upon the measurement
of the invariant mass spectra of $e^{+}e^{-}$ pairs in 12 GeV $p+A$ reactions,
with copper and carbon being used as targets~\cite{Muto:2005za}. Even though this result
may indicate a signal for partial restoration of chiral symmetry in nuclear
matter, it is not possible to draw a definite conclusion solely from this.
In fact, recently, a large in-medium $\phi$-meson decay width ($>30$ MeV)
has been extracted at various experimental facilities without observing any
mass shift~\cite{Ishikawa:2004id,Mibe:2007aa,Qian:2009ab,Wood:2010ei,Polyanskiy:2010tj}. It is therefore clear that the search for evidence of
a light vector meson mass shift in nuclear matter is indeed a complicated
issue and further experimental efforts~\cite{Aoki:2015qla,JPARCE16Proposal} are required in order to
understand the phenomenon better. Indeed, the J-PARC E16 collaboration~\cite{Aoki:2015qla,JPARCE16Proposal}
intends to perform a more systematic study for the mass shift of vector mesons
with higher statistics than the above-mentioned experiment at KEK-E325.
However, either complementary or alternative experimental methods are desired.
The study of the $\phi$-meson--nucleus bound states is complementary to the
invariant mass measurements, such as Ref.~\cite{Muto:2005za}, where only a small fraction
of the produced $\phi$-mesons decay inside the nucleus and may be expected to
provide extra information on the $\phi$-meson properties at finite baryon
density. Along these lines, and motivated by the 3.4\% mass reduction reported
by the KEK-E325 experiment~\cite{Muto:2005za}, the E29 collaboration at J-PARC has recently put forward a proposal~\cite{JPARCE29Proposal,JPARCE29ProposalAdd} to study the in-medium mass modification of
the $\phi$-meson via the possible formation of $\phi$-meson--nucleus bound
states~\cite{Ohnishi:2014xla,Buhler:2010zz}. Furthermore, there is also a proposal at JLab, following the
12 GeV upgrade, to study the binding of $\phi$ and $\eta$ mesons to
$^{4}$He~\cite{JLabphiJLabphi}. This new experimental approach~\cite{Ohnishi:2014xla,Aoki:2015qla,Buhler:2010zz,JLabphiJLabphi} for the
measurement of the $\phi$-meson mass shift in nuclei, will produce a slowly
moving $\phi$-meson~\cite{Ohnishi:2014xla,Aoki:2015qla,Buhler:2010zz,JLabphiJLabphi}, where the maximum nuclear matter effect
can be probed. In this way, one may indeed anticipate the formation of a
$\phi$-meson--nucleus bound state, where the $\phi$-meson is trapped inside
the nucleus. Meson-nucleus systems bound by attractive strong interactions
are very interesting objects~\cite{Krein:2017usp,Metag:2017yuh}. First, they are strongly interacting exotic
many-body systems and to study them serves, for example, to understand better
the multiple-gluon exchange interactions, including ``QCD van der Waals''
forces~\cite{Appelquist:1978rt}, which are believed to play a role in the binding of the $J/\Psi$
and other exotic heavy-quarkonia to matter (a nucleus)~\cite{Brodsky:1989jd,Luke:1992tm,Sibirtsev:1999jr,Gao:2000az,Beane:2014sda,Brambilla:2015rqa,Gao:2017hya,Kawama:2014pja,Kawama:2014iwa,Ohnishi:2014xla,Aoki:2015qla,Morino:2015bqa}. Second, they
provide unique laboratories for the study of hadron properties at finite
density, which may not only lead to a deeper understanding of the strong
interaction~\cite{Hayano:2008vn,Leupold:2009kz,Hosaka:2016ypm,Krein:2017usp,Metag:2017yuh} but of the structure of finite nuclei as well~\cite{Stone:2016qmi,Guichon:2006er}.
A downward mass shift of the $\phi$-meson in a nucleus is directly connected
with the possible existence of an attractive potential between the $\phi$-meson
and the nucleus where it has been produced, the strength of which is expected
to be of the same order as that of the mass shift. Concerning the theoretical
evaluation of the $\phi$-meson mass shift, various authors predict a downward
shift of the in-medium $\phi$-meson mass and a broadening of its decay width,
many of them focusing on the self-energy of the $\phi$-meson due to the
kaon-antikaon loop. Ko et al.~\cite{Ko:1992tp} used a density-dependent kaon mass
determined from chiral perturbation theory and found that at normal nuclear
matter density, $\rho_{0}$, the $\phi$-meson mass decreases very little,
by at most 2\%, and a $\Gamma_{\phi}\approx 25$ MeV width which broadens
drastically for large densities. Hatsuda and Lee calculated the in-medium
$\phi$-meson mass based on the QCD sum rule approach~\cite{Hatsuda:1991ez,Hatsuda:1996xt}, and predict a
1.5\%-3\% decrease at $\rho_{0}$. Other investigations also predict a small
downward mass shift and a large broadening of the $\phi$-meson width at
$\rho_{0}$: Ref.~\cite{Klingl:1997tm} reports a negative mass shift of $< 1\%$ and a decay
width of 45 MeV; Ref.~\cite{Oset:2000eg} predicts a decay width of 22 MeV but does not
report a result on the mass shift; and Ref.~\cite{Cabrera:2002hc} gives a rather small
negative mass shift of $\approx 0.81\%$ and a decay width of 30 MeV.
More recently, Ref.~\cite{Gubler:2015yna} reported a downward mass shift of $< 2\%$ and a
large broadening width of 45 MeV at $\rho_{0}$; and finally, in Ref.~\cite{Cabrera:2016rnc},
extending the work of Refs.~\cite{Oset:2000eg,Cabrera:2002hc}, the authors reported a negative mass
shift of 3.4\% and a large decay width of 70 MeV at $\rho_{0}$. The reason
for these differences may lie in the different approaches used to estimate
the kaon-antikaon loop contributions to the $\phi$-meson self-energy and
this might have consequences for the formation of $\phi$-meson--nucleus bound
states. From a practical point of view, the important question is whether
this attraction, if it exists, is sufficient to bind the $\phi$-meson to a
nucleus. A simple argument can be given as follows. One knows that for an
attractive spherical well of radius $R$ and depth $V_{0}$, the condition for
the existence of a nonrelativistic $s$-wave bound state of a particle of mass
$m$ is $V_{0}>\frac{\pi^{2}\hbar^{2}}{8mR^{2}}$. Using $m = m_{\phi}^{*}$, where
$m_{\phi}^{*}$ is the $\phi$-meson mass at normal nuclear matter density found
in Ref.~\cite{Muto:2005za} and $R = 5$ fm (the radius of a heavy nucleus), one obtains
$V_{0}>2$ MeV. Therefore, the prospects of capturing a $\phi$-meson seems
quite favorable, provided that the $\phi$-meson can be produced almost at
rest in the nucleus. Real nuclei, however, have a surface and this estimate
can be quite misleading~\cite{Krein:2017usp}. A full calculation using realistic density
profiles of nuclei is required for a more reliable estimate. The work
presented at this workshop was carried out in collaboration with
professors K.~Tushima, G.~Krein, and A.~W.~Thomas and has been published in
Refs.~\cite{Cobos-Martinez:2017vtr,Cobos-Martinez:2017woo}. In Ref.~\cite{Cobos-Martinez:2017vtr} we studied the $\phi$-meson mass shift and
decay width in nuclear matter, based on an effective Lagrangian approach, by
evaluating the $K\overline{K}$ loop contribution in the $\phi$-meson
self-energy, with the in-medium K and $\overline{K}$ masses explicitly
calculated by the quark-meson coupling (QMC) model~\cite{Saito:2005rv}. This initial study
has been extended in Ref.~\cite{Cobos-Martinez:2017woo} to some selected nuclei by computing the
$\phi$-meson--nucleus bound complex potential in the local density
approximation and solving the Klein-Gordon equation in order to obtain the
bound state energies and absorption widths. The nuclear density distributions
for all nuclei studied, except for $^{4}$He, are explicitly calculated using the
QMC model~\cite{Saito:1996sf}.

\section{$\phi$-meson mass and decay width in nuclear matter}

\begin{figure}
  \centering
\scalebox{0.9}{
\begin{tabular}{ccc}
 \includegraphics[scale=0.2]{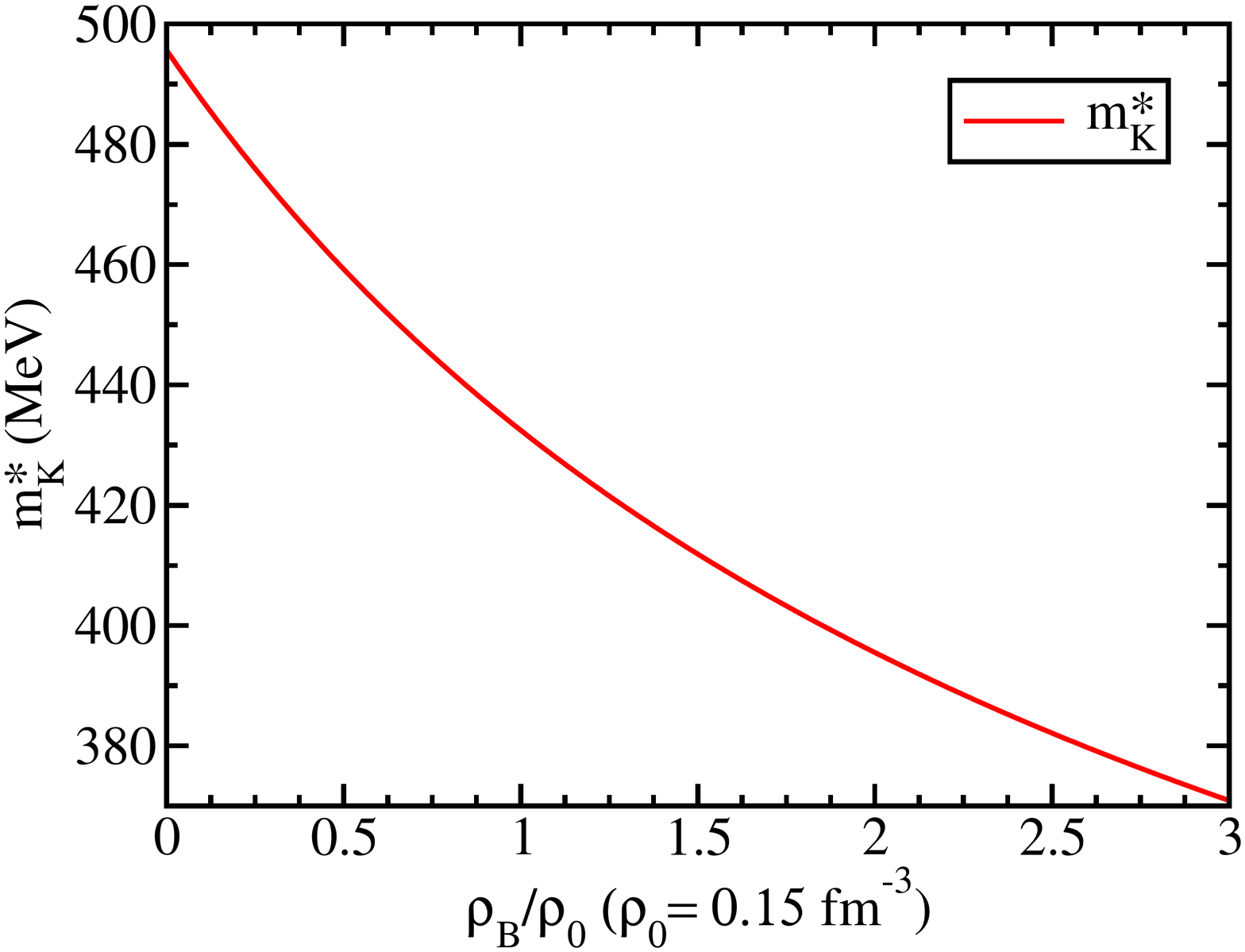} &
 \includegraphics[scale=0.2]{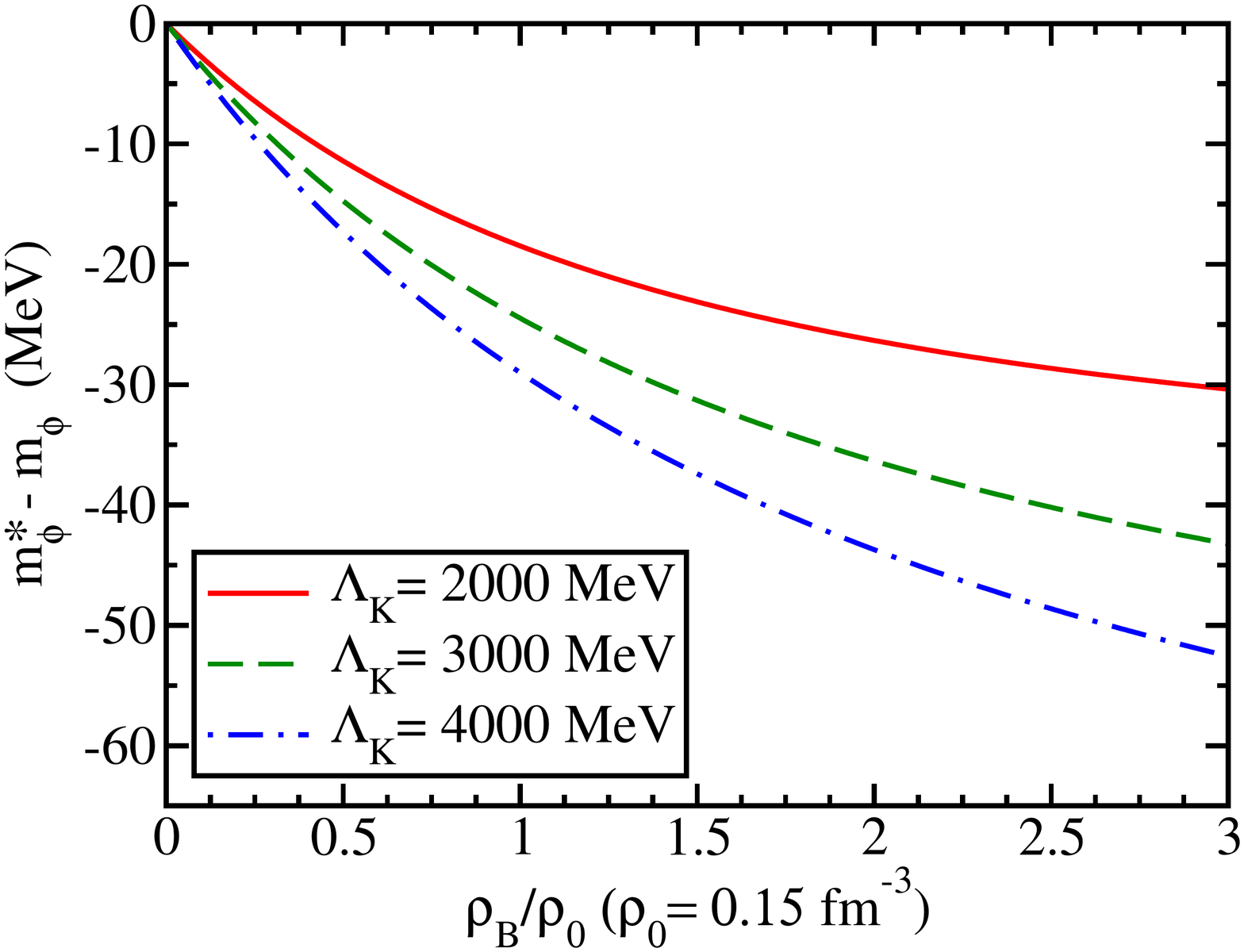} &
 \includegraphics[scale=0.176]{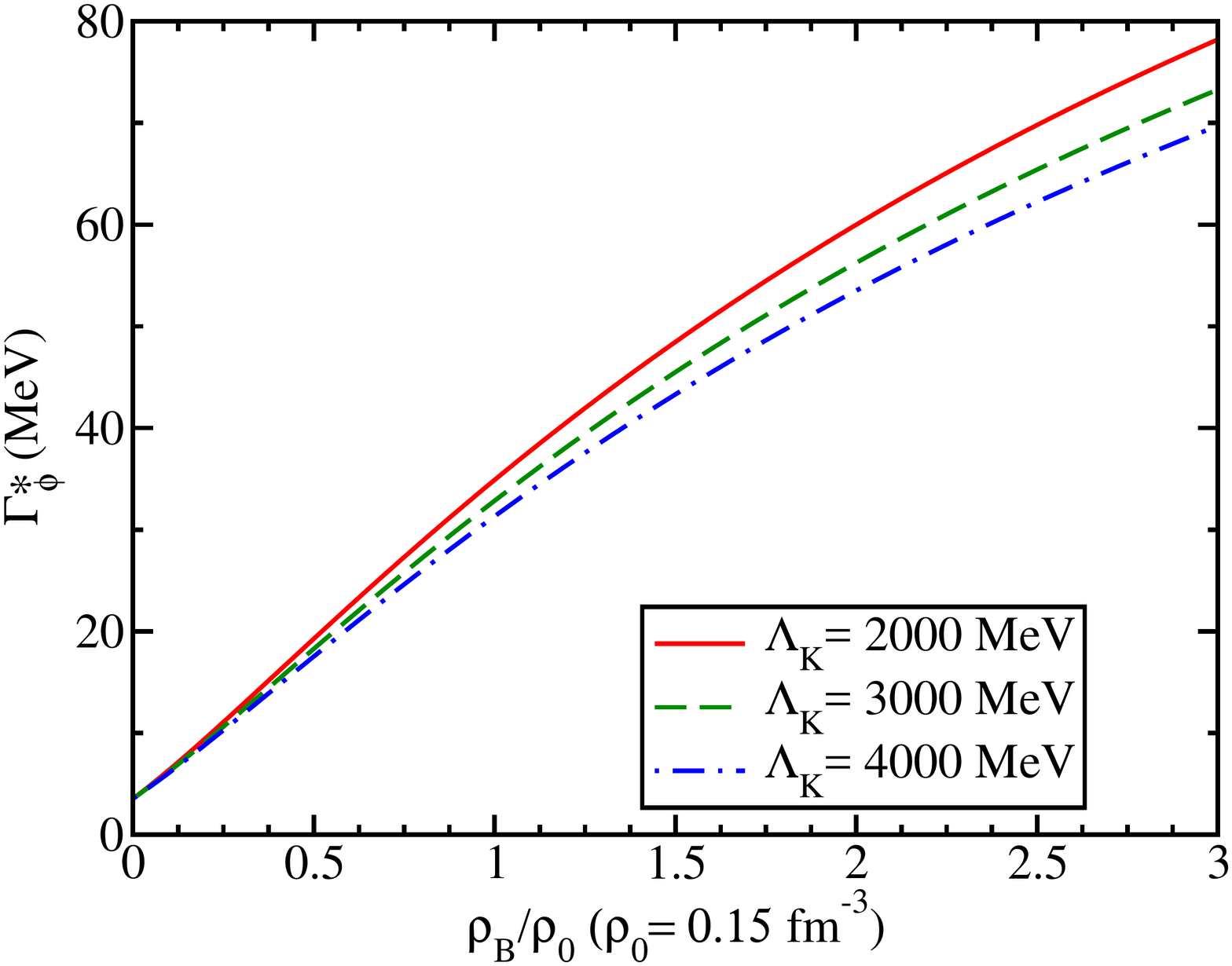} \\
\end{tabular}
}
\caption{\label{fig:nuclmatt} Left panel: In-medium kaon mass (=antikaon)
  Lorentz scalar mass $m_{K}^{*}$; centre and right panels: $\phi$-meson mass
  shift and decay width, respectively, in symmetric nuclear matter for three
  values of the cutoff parameter $\Lambda_{K}$.}
\end{figure}

The $\phi$-meson property modifications in nuclear matter, such as its mass
and decay width, are strongly correlated to its coupling to the
$K\overline{K}$ channel, which is the dominant decay channel in vacuum.
Thus, one expects that a significant fraction of the density dependence of
the $\phi$-meson self-energy in nuclear matter arises from the in-medium
modification of the KK intermediate state in the $\phi$-meson self-energy.
We briefly review the computation~\cite{Cobos-Martinez:2017vtr} of the $\phi$-meson self-energy in
vacuum and in nuclear matter using an effective Lagrangian approach~\cite{Klingl:1996by}.
The interaction Lagrangian $\mathcal{L}_{\text{int}}$ involves
$\phi K\overline{K}$ and $\phi\phi K\overline{K}$ couplings dictated by a
local gauge symmetry principle:
\begin{equation}
  \label{eqn:Lint}
  \mathcal{L}_{\text{int}}=  \mathcal{L}_{\phi K\overline{K}}
  +\mathcal{L}_{\phi\phi K\overline{K}}
  \end{equation}
\noindent where (we use the convention
$K=\left(\begin{array}{c} K^{+} \\ K^{0} \end{array}\right)$ for isospin
doublets)
\begin{eqnarray}
\label{eqn:Lpkk}
\mathcal{L}_{\phi K\overline{K}}&=&\mi g_{\phi}\phi^{\mu}
\left[\Kbar(\partial_{\mu}K)-(\partial_{\mu}\Kbar)K\right], \\
\label{eqn:Lppkk}
\mathcal{L}_{\phi\phi K\Kbar}&=& g^2_{\phi} \phi^\mu\phi_\mu \Kbar K.
\end{eqnarray}
We note that the use of the effective interaction Lagrangian of \eqn{eqn:Lint}
without the term given in \eqn{eqn:Lppkk} may be considered as being motivated
by the hidden gauge approach in which there are no four-point vertices,
such as \eqn{eqn:Lppkk}, that involve two pseudoscalar mesons and two vector
mesons~\cite{Lin:1999ve,Lee:1994wx}. This is in contrast to the approach of using the minimal
substitution to introduce vector mesons as gauge particles where such
four-point vertices do appear. However, these two methods have been shown to
be consistent if both the vector and axial vector mesons are included~\cite{Yamawaki:1986zz,Meissner:1986tc,Meissner:1987ge,Saito:1987ba}.
Therefore, we present results with and without such an interaction~\cite{Cobos-Martinez:2017vtr,Cobos-Martinez:2017woo}.
We consider first the contribution from the $\phi K\overline{K}$ coupling,
given by \eqn{eqn:Lpkk}, to the scalar part of the $\phi$-meson self-energy,
$\Pi_{\phi}(p)$.
\begin{equation}
\label{eqn:phise}
\mi\Pi_{\phi}(p)=-\frac{8}{3}g_{\phi}^{2}\int\dfd{4}{4}{q}\vec{q}^{\,2}
D_{K}(q)D_{K}(q-p) \, ,
\end{equation}
\noindent where $D_{K}(q)=\left(q^{2}-m_{K}^{2}+\mi\epsilon\right)^{-1}$ is the
kaon propagator;  $p=(p^{0}=m_{\phi},\vec{0})$ is the $\phi$ meson four-momentum
vector at rest, with $m_{\phi}$ the $\phi$ meson mass; $m_{K} (=m_{\Kbar})$ is
the kaon mass, and $g_{\phi}$ is the coupling constant. When $m_{\phi}<2m_{K}$
the self-energy $\Pi_{\phi}(p)$ is real. However, when  $m_{\phi}>2m_{K}$, which
is the case here, $\Pi_{\phi}(p)$ acquires an imaginary part. The integral in
\eqn{eqn:phise} is divergent but it will be regulated using a phenomenological
form factor, with cutoff parameter $\Lambda_{K}$, as in Refs.~\cite{Cobos-Martinez:2017vtr,Krein:2010vp}.
The sensitivity of the results to the cutoff value is analyzed below. The mass and
decay width of the $\phi$-meson in vacuum ($m_{\phi}$ and $\Gamma_{\phi}$), as
well as in nuclear matter ($m_{\phi}^{*}$ and $\Gamma_{\phi}^{*}$), are
determined self-consistently~\cite{Cobos-Martinez:2017vtr} from
\begin{eqnarray}
\label{eqn:phimass} 
m_{\phi}^{2}&=&\left(m_{\phi}^{0}\right)^{2}+\Re\Pi_{\phi}(m_{\phi}^{2}), \\
\label{eqn:phiwidth}
\Gamma_{\phi}&=&-\frac{1}{m_{\phi}}\Im\Pi_{\phi}(m_{\phi}^{2}).
\end{eqnarray}
The coupling constant $g_{\phi}$ is determined~\cite{Cobos-Martinez:2017vtr} from \eqns{eqn:phimass}
and~(\ref{eqn:phiwidth}), and the experimental value for the $\phi\to K\Kbar$
decay width in vacuum, corresponding to the branching ratio of 83.1\% of the
total decay width~\cite{PDG:2015}.
The nuclear density dependence of the $\phi$-meson mass and decay width is
driven by the intermediate $K\Kbar$ state interactions with the nuclear medium.
This effect enters through $m_{K}^{*}$ in the kaon propagators in
\eqn{eqn:phise}. The in-medium mass $m_{K}^{*}$is calculated within the QMC
model~\cite{Cobos-Martinez:2017vtr}, which has proven to be very successful in studying the properties
of hadrons in nuclear matter and finite nuclei. For a more complete discussion
of the model see Refs.~\cite{Tsushima:1997df,Saito:2005rv,Guichon:1989tx}. Here we just make a few necessary comments.
In order to calculate the in-medium properties of $K$ and $\Kbar$, we consider
infinitely large, uniformly symmetric, spin-isospin-saturated nuclear matter
in its rest frame, where all the scalar and vector mean field potentials,
which are responsible for the nuclear many-body interactions, become constant
in the Hartree approximation~\cite{Cobos-Martinez:2017vtr}.
In \fig{fig:nuclmatt} (left panel) we present the resulting in-medium kaon
Lorentz scalar mass (=antikaon Lorentz scalar mass), calculated using the QMC
model, as a function of the baryon density. The kaon effective mass at normal
nuclear matter density $\rho_{0}= 0.15$ fm$^{-3}$ has decreased by about 13\%.
We also recall, in connection with the calculation of the in-medium $K\Kbar$
loop contributions to the $\phi$-meson self-energy, that the isoscalar-vector
$\omega$ mean field potentials arise both for the kaon and antikaon. However,
they have opposite signs and cancel each other. Equivalently, they can be
eliminated by a variable shift in the loop calculation~\cite{Tsushima:1997df,Saito:2005rv,Guichon:1989tx} of the
$\phi$-meson self-energy, and therefore we do not show them here.
In \fig{fig:nuclmatt}, we present the $\phi$-meson mass shift (centre panel)
and decay width (right panel) as a function of the nuclear matter density,
$\rho_{B}$, for three values of the cutoff parameter $\Lambda_{K}$. As can be
seen, the effect of the in-medium kaon and antikaon mass change yields a
negative mass shift for the $\phi$-meson. This is because the reduction in the
kaon and antikaon masses enhances the $K\Kbar$ loop contribution in nuclear
matter relative to that in vacuum. For the largest value of the nuclear matter
density, the downward mass shift turns out to be a few percent at most for
all values of $\Lambda_{K}$. On the other hand, we see that $\Gamma_{\phi}^{*}$
is very sensitive to the change in the kaon and antikaon masses, increasing
rapidly with increasing nuclear matter density, up to a factor of $\approx 20$
enhancement for the largest value of $\rho_{B}$. In \tab{tab:nuclmatt} we
present the values for $m_{\phi}^{*}$and $\Gamma_{\phi}^{*}$ at normal nuclear
matter density $\rho_{0}$. We see that the negative kaon and antikaon mass
shift of 13\%~\cite{Cobos-Martinez:2017vtr} induces a downward mass shift of the $\phi$-meson of just
$\approx 2\%$, while the broadening of the $\phi$-meson decay width is an
order-of-magnitude larger than its vacuum value.
\begin{table}
  \centering
\scalebox{1.0}{
  \begin{tabular}{c|ccc}
    \hline
    & $\Lambda_{K}=2000$ & $\Lambda_{K}= 3000$ & $\Lambda_{K}= 4000$ \\
    \hline
$m_{\phi}^{*}$ & 1000.9 (1009.5) & 994.9 (1004.3) & 990.4 (1000.6) \\
$\Gamma_{\phi}^{*}$ & 34.8 (37.8) & 32.8 (36.0) & 31.3 (34.7) \\
  \hline
\end{tabular}
}
\caption{\label{tab:nuclmatt} $\phi$-meson mass and width at normal nuclear
  matter density $\rho_{0}$ with and without the gauged Lagrangian of
  \eqn{eqn:Lppkk}. The values in parentheses were computed by adding the
  gauged Lagrangian of \eqn{eqn:Lppkk}. All quantities are given in MeV.}
\end{table}
In Ref.~\cite{Cobos-Martinez:2017vtr} we evaluated the impact of adding the $\phi\phi K\Kbar$
interaction of \eqn{eqn:Lppkk} on the in-medium $\phi$-meson mass and
decay width. We found that one still gets a downward shift of the in-medium
$\phi$-meson mass as well as a significant broadening of the decay width when
this interaction is added. In both cases, though, the absolute values are
slightly different from those shown in \fig{fig:nuclmatt}.
In tab{tab:nuclmatt} we present the values for $m_{\phi}^{*}$ and
$\Gamma_{\phi}^{*}$ at $\rho_{0}$ obtained by adding the gauged Lagrangian of
\eqn{eqn:Lppkk}. In both cases, for the mass and decay width in nuclear
matter, the effect of adding \eqn{eqn:Lppkk} can be compensated by the use
of a larger cutoff $\Lambda_{K}$.
The results described above support those which suggest that one should
observe a small downward mass shift and a large broadening of the decay width
of the $\phi$-meson in a nuclear medium. Furthermore, they open experimental
possibilities for studying the binding and absorption of $\phi$-meson in
nuclei. Although the mass shift found in this study may be large enough to
bind the $\phi$-meson to a nucleus, the broadening of its decay width will
make it difficult to observe a signal for the $\phi$-meson--nucleus bound
state formation experimentally. We explore this further in the second part of
this talk.

\section{$\phi$-meson--nucleus bound states }

In this part we discuss the situation where the $\phi$-meson is placed in a
nucleus. The nuclear density distributions for the nuclei $^{12}$C, $^{16}$O,
and $^{208}$Pb are obtained using the QMC model~\cite{Saito:1996sf}. For $^{4}$He, we use the
parametrization for the density distribution obtained in Ref.~\cite{Saito:1997ae}. Then,
using a local density approximation we calculate the $\phi$-meson complex
potentials for a nucleus $A$, which can be written as
\begin{equation}
\label{eqn:Vcomplex}
V_{\phi A}(r)= U_{\phi}(r)-\frac{\mi}{2}W_{\phi}(r),
\end{equation}
\noindent where $r$ is the distance from the center of the nucleus and 
$U_{\phi}(r)=\Delta m_{\phi}(\rho_{B}(r)) \equiv m^{*}_{\phi}(\rho_{B}(r))-m_\phi$ 
and $W_{\phi}(r)=\Gamma_{\phi}(\rho_{B}(r))$ are, respectively,  
the $\phi$-meson mass shift and decay width in a nucleus $A$, with 
$\rho_{B}(r)$ the baryon density distribution for the particular nucleus.
\begin{figure}
  \centering
\scalebox{0.675}{
\begin{tabular}{cccc}
  \includegraphics[scale=0.2]{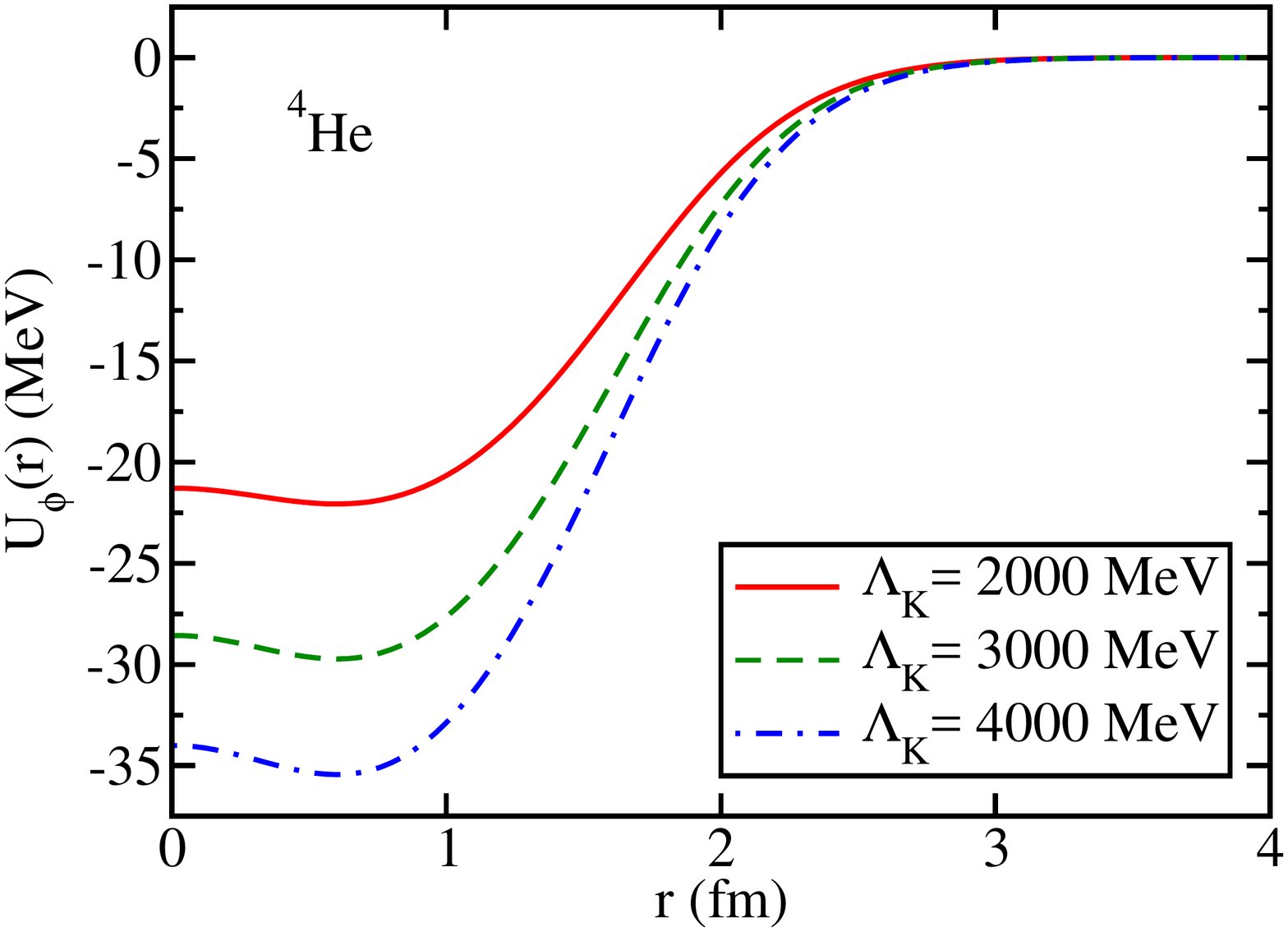} &
  \includegraphics[scale=0.2]{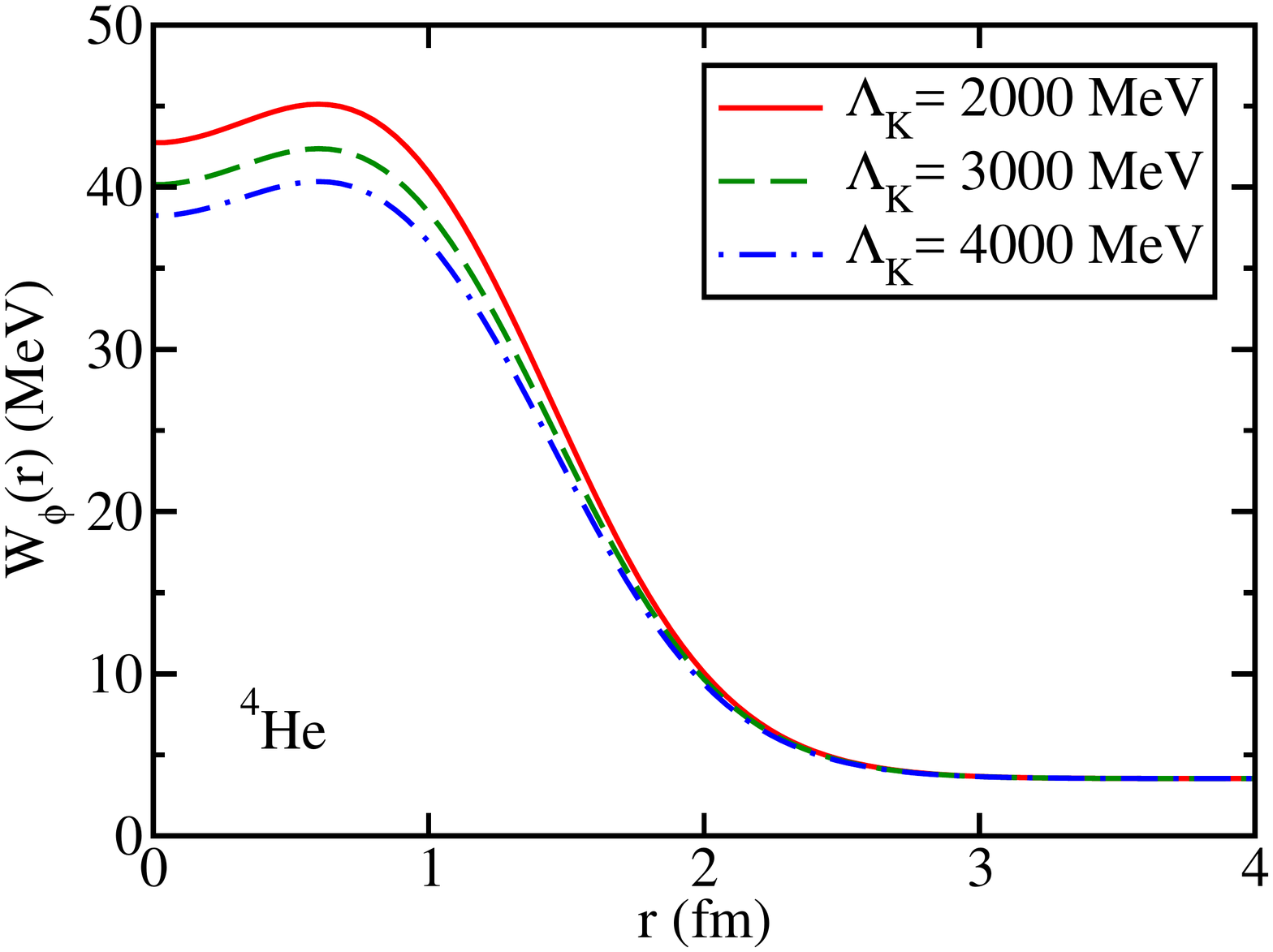} &
  \includegraphics[scale=0.2]{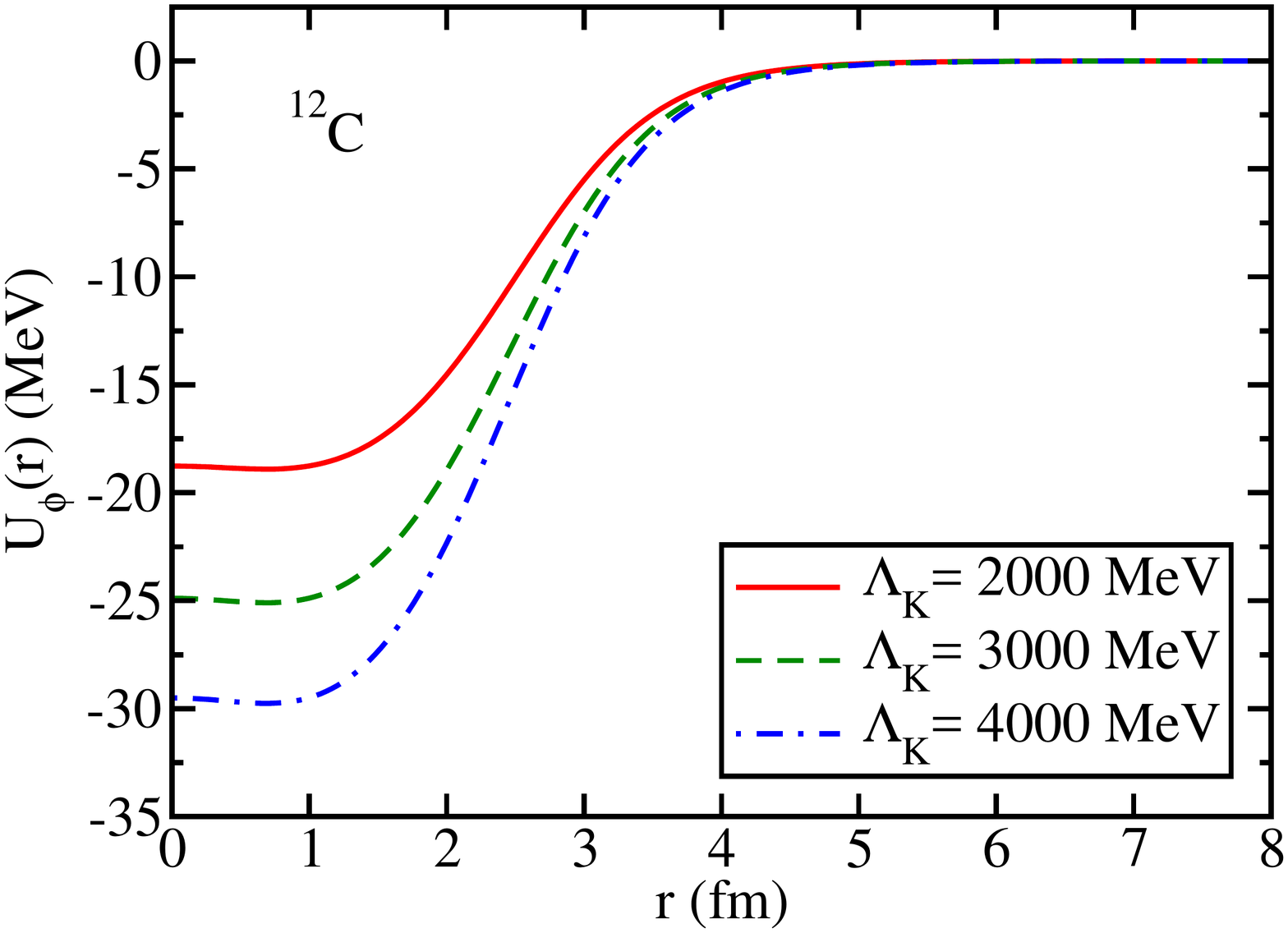} &
  \includegraphics[scale=0.2]{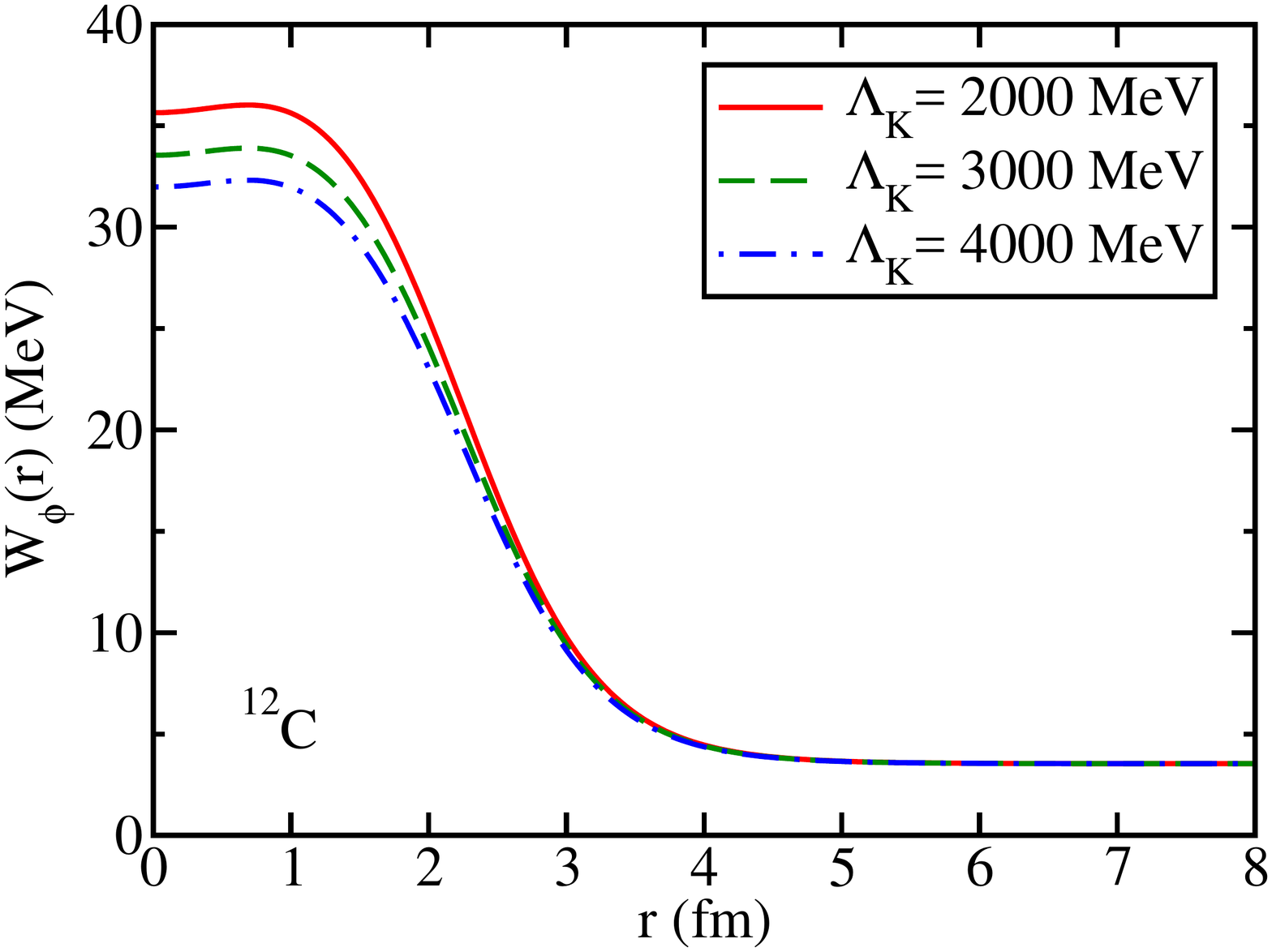} \\
  \includegraphics[scale=0.2]{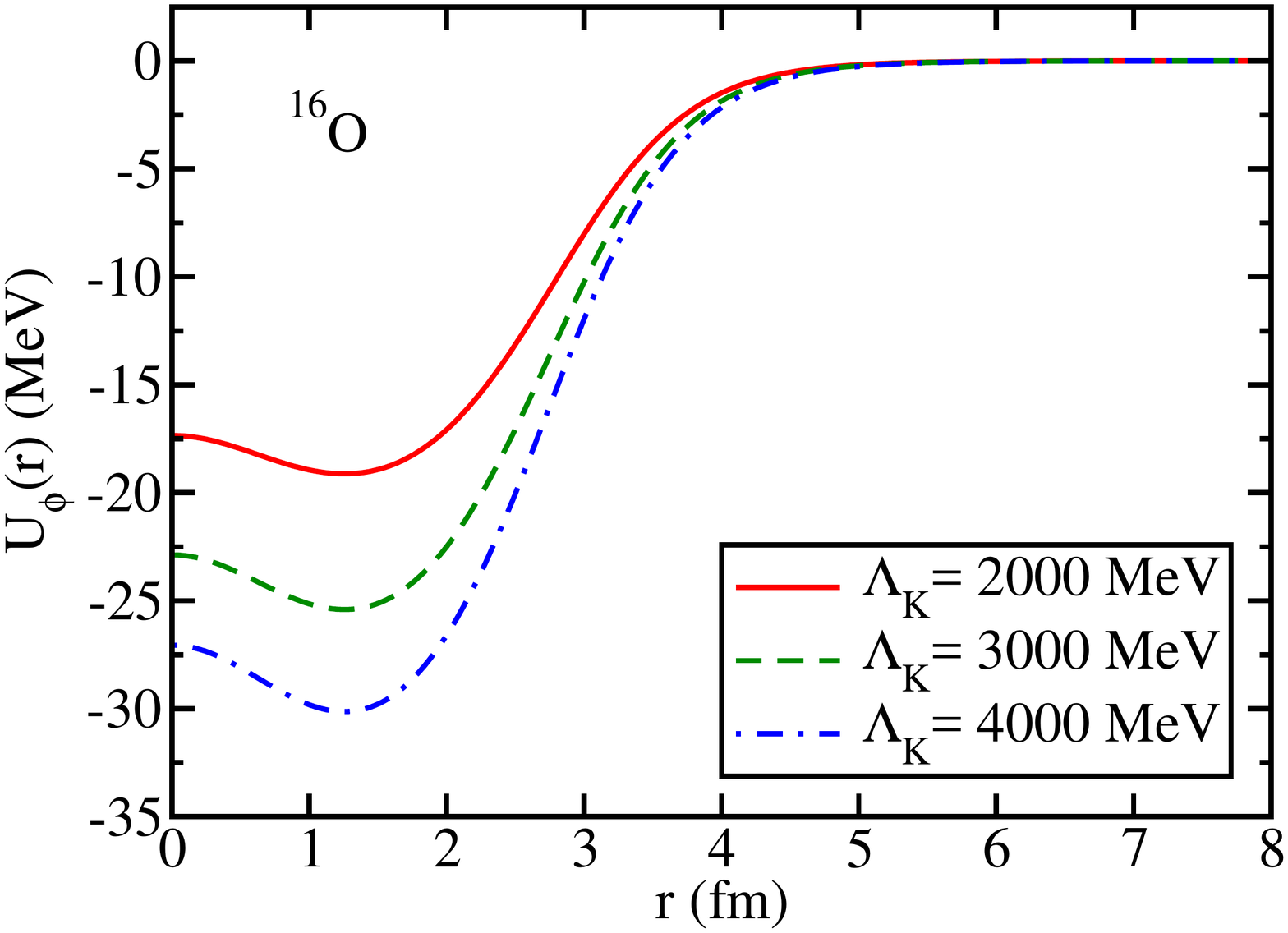} &
  \includegraphics[scale=0.2]{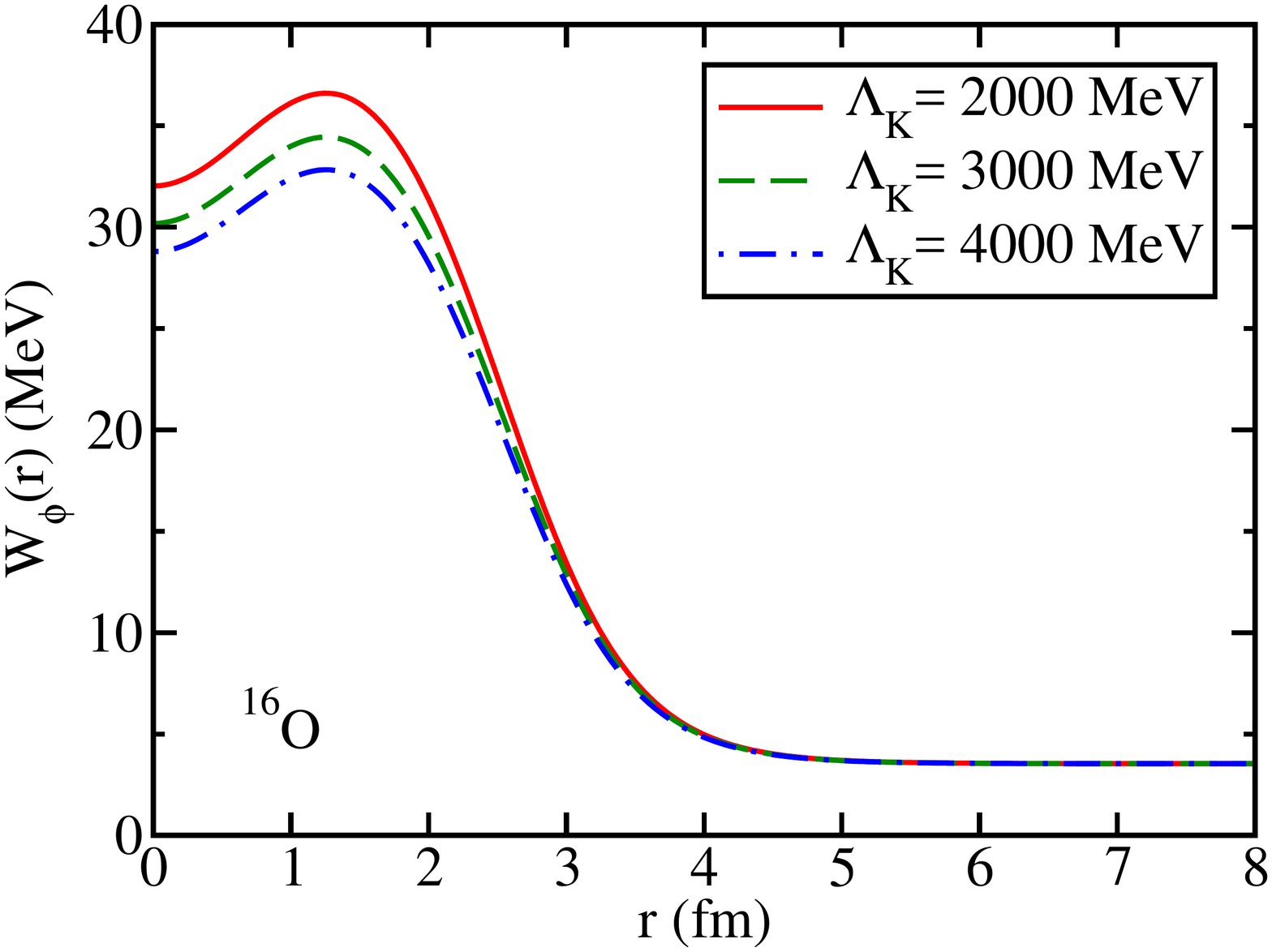} &
  \includegraphics[scale=0.2]{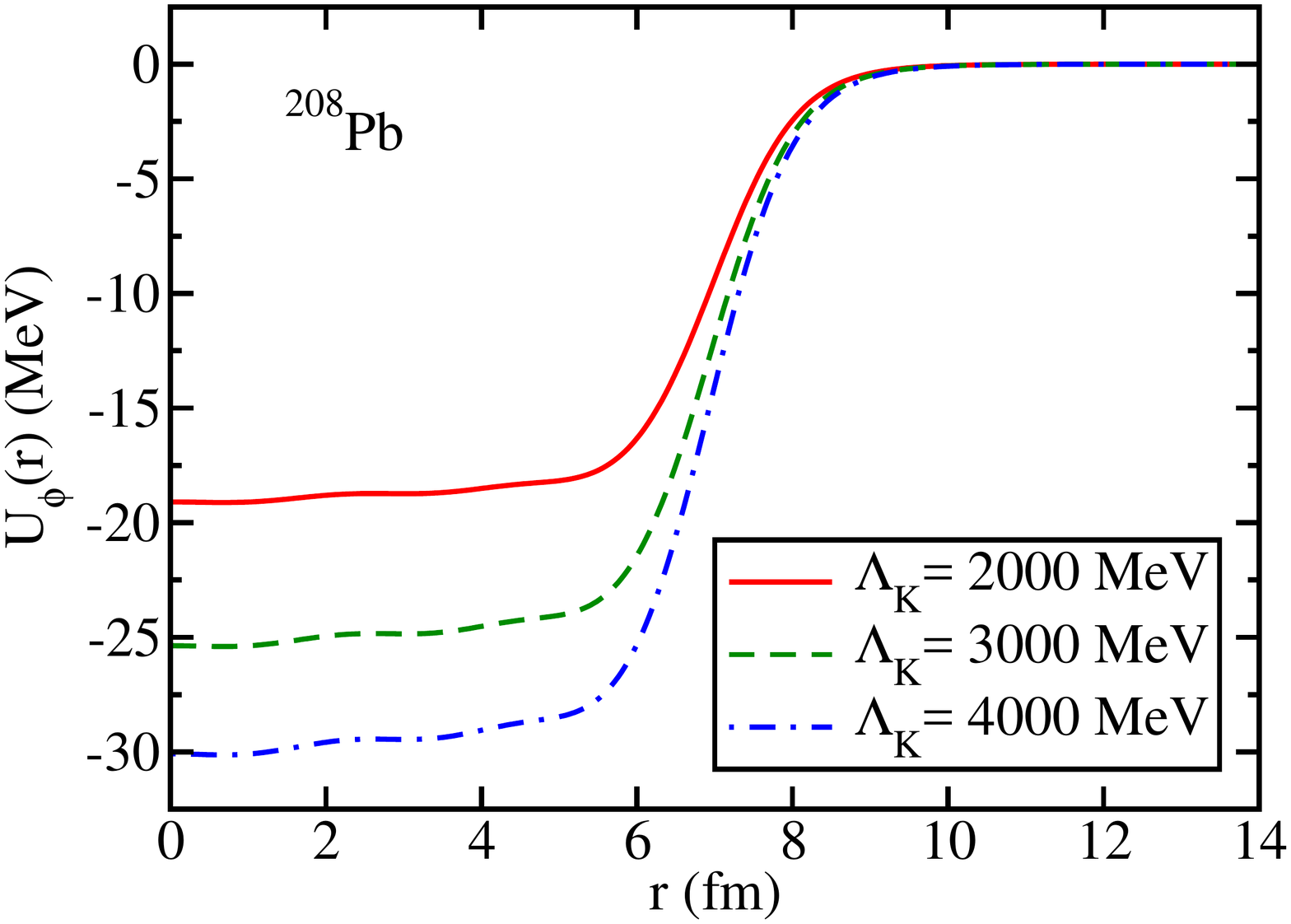} &
  \includegraphics[scale=0.2]{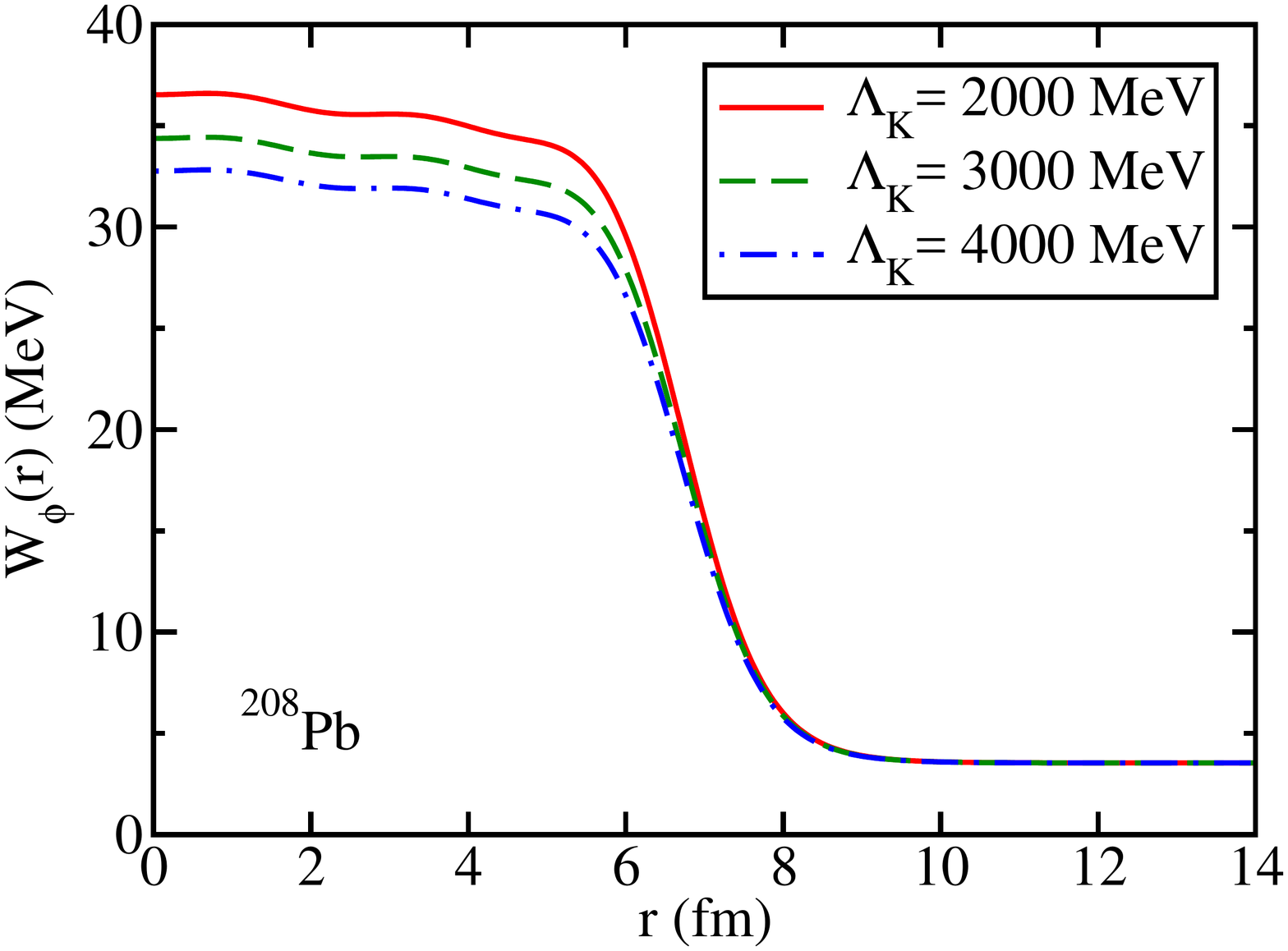} \\
\end{tabular}
}
\caption{\label{fig:phinuclpot} Real [$U_{\phi}(r)$] and imaginary [$W\phi(r)$]
  parts of the $\phi$-meson--nucleus potentials for the nuclei $^{4}$He,
  $^{12}$C, $^{16}$O, and $^{208}$Pb, for three values of the cutoff parameter
  $\Lambda_{K}$.}
\end{figure}
\fig{fig:phinuclpot} shows the $\phi$-meson potentials calculated for the
selected nuclei, for three values of the cutoff parameter $\Lambda_{K}$. One
can see that the depth of the real part of the potential is sensitive to the
cutoff parameter, varying from -20 MeV to -35 MeV for $^{4}$He and from
-20 MeV to -30 MeV for $^{208}$Pb. On the other hand, one can see that the
imaginary part does not vary much with $\Lambda_{K}$. These latter observations
may well have consequences for the feasibility of experimental observation
of the expected bound states.
Using the $\phi$-meson potentials obtained in this manner, we next calculate
the $\phi$-meson--nucleus bound state energies and absorption widths for the
selected nuclei. Before proceeding, a few comments on the use of \eqn{eqn:kge}
are in order. In this study, we consider the situation where the $\phi$-meson
is produced nearly at rest. Then, it should be a very good approximation to
neglect the possible energy difference between the longitudinal and transverse
components of the $\phi$-meson wave function $\psi_{\phi}^{\mu}$. After imposing
the Lorentz condition, $\partial_{\mu}\psi_{\phi}^{\mu}=0$, to solve the Proca
equation becomes equivalent to solving the Klein-Gordon equation
\begin{equation}
\label{eqn:kge}
\left(-\nabla^{2} + \mu^{2} + 2\mu V(\vec{r})\right)\phi(\vec{r})
= \mathcal{E}^{2}\phi(\vec{r}),
\end{equation}
where $μ = m_{\phi}m_{A}/(m_{\phi} + m_{A})$ is the reduced mass of the
$\phi$-meson--nucleus system with $m_{\phi}$ ($m_{A}$) the mass of the
$\phi$-meson (nucleus $A$) in vacuum, and $V (r)$ is the complex
$\phi$-meson--nucleus potential of \eqn{eqn:Vcomplex}. We solve the
Klein-Gordon equation using the momentum space methods~\cite{Kwan:1978zh}.
The calculated bound state energies ($E$) and absorption widths ($\Gamma$), which are
related to the complex energy eigenvalue $\mathcal{E}$ by
$E= \Re\mathcal{E}-\mu$ and $\Gamma= -2\Im\mathcal{E}$, are listed in
\tab{tab:phienergies} for three values of the cutoff parameter $\Lambda_{K}$,
with and without the imaginary part of the potential.
We first discuss the case in which the imaginary part of the
$\phi$-meson--nucleus potential is set to zero. The results are given within
parentheses in \tab{tab:phienergies}. From the values shown, we see that the
$\phi$-meson is expected to form bound states with all the nuclei selected,
for all values of the cutoff parameter $\Lambda_{K}$. The bound state energy
is obviously dependent on $\Lambda_{K}$, increasing as $\Lambda_{K}$ increases.
Next, we discuss the results obtained when the imaginary part of the potential
is included. Adding the absorptive part of the potential changes the situation
appreciably. From the results presented in \tab{tab:phienergies} we note that
for the largest value of the cutoff parameter $\Lambda_{K}$, which yields the
deepest attractive potentials (see \fig{fig:phinuclpot}), the $\phi$-meson
is expected to form bound states with all the selected nuclei, including the
lightest one, the $^{4}$He nucleus. However, in this case, whether or not the
bound states can be observed experimentally is sensitive to the value of the
cutoff parameter $\Lambda_{K}$. One also observes that the width of the bound
state is insensitive to the values of $\Lambda_{K}$ for all nuclei.
Furthermore, since the so-called dispersive effect of the absorptive potential
is repulsive, the bound states disappear completely in some cases, even though
they were found when the absorptive part was set to zero. This feature is
obvious for the $^{4}$He nucleus, making it especially relevant to the future
experiments, planned at J-PARC and JLab using light and medium-heavy
nuclei~\cite{Ohnishi:2014xla,Aoki:2015qla,Buhler:2010zz,JLabphiJLabphi}.
%
\begin{table}[ht]
\begin{center}
\scalebox{0.85}{
  \begin{tabular}{ll|rr|rr|rr} 
\hline \hline
& & \multicolumn{2}{c|}{$\Lambda_{K}=2000$} &
\multicolumn{2}{c}{$\Lambda_{K}=3000$} & 
\multicolumn{2}{c}{$\Lambda_{K}=4000$}  \\
\hline
 & & $E$ & $\Gamma/2$ & $E$ & $\Gamma/2$ & $E$ & $\Gamma/2$ \\
\hline
$^{4}_{\phi}\text{He}$ & 1s & n (-0.8) & n & n (-1.4) & n & -1.0 (-3.2) & 8.3 \\
\hline
$^{12}_{\phi}\text{C}$ & 1s & -2.1 (-4.2) & 10.6 & -6.4 (-7.7) & 11.1 & -9.8
(-10.7) & 11.2 \\
\hline
$^{16}_{\phi}\text{O}$ & 1s & -4.0 (-5.9) & 12.3 & -8.9 (-10.0) & 12.5 & -12.6
(-13.4) & 12.4 \\
& 1p & n (n) & n & n (n) & n & n (-1.5) & n \\
\hline
$^{208}_{\phi}\text{Pb}$ & 1s & -15.0 (-15.5) & 17.4 & -21.1 (-21.4) & 16.6 &
-25.8 (-26.0) & 16.0 \\
& 1p & -11.4 (-12.1) & 16.7 & -17.4 (-17.8) & 16.0 & -21.9  (-22.2) & 15.5 \\
& 1d & -6.9 (-8.1) & 15.7 & -12.7 (-13.4) & 15.2 & -17.1 (-17.6) & 14.8 \\
& 2s & -5.2 (-6.6) & 15.1 & -10.9 (-11.7) & 14.8 & -15.2 (-15.8) & 14.5 \\
& 2p & n (-1.9) & n & -4.8 (-6.1) & 13.5 & -8.9 (-9.8) & 13.4 \\
& 2d & n (n) & n & n (-0.7) & n & -2.2 (-3.7) & 11.9 \\
\hline \hline
\end{tabular}}
\caption{\label{tab:phienergies} $\phi$-nucleus single-particle energies $E$
  and half width $\Gamma/2$ obtained, with and without the imaginary part of
  the potential of \eqn{eqn:Vcomplex}, for three values of the cutoff parameter
  $\Lambda_K$. When only the real part is included, the corresponding
  single-particle energy $E$ is given inside parenthesis and $\Gamma=0$ for
  all nuclei. ``n'' indicates that no bound state is found. All quantities are
  given in MeV.}
\end{center}
\end{table}

\section{Summary and discussion}

We have calculated the $\phi$-meson mass and width in nuclear matter within an
effective Lagrangian approach up to three times of normal nuclear matter
density. Essential to our results are the in-medium kaon masses, which are
calculated in the QMC model, where the scalar and vector meson mean fields
couple directly to the light $u$ and $d$ quarks (antiquarks) in the $K$
($\Kbar$) mesons. At normal nuclear matter density, allowing for a very large
variation of the cutoff parameter $\Lambda_{K}$, although we have found a
sizable negative mass shift of 13\% in the kaon mass, this induces only a
few percent downward shift of the $\phi$-meson mass. On the other hand, it
induces an order-of-magnitude broadening of the decay width. We have also
calculated the $\phi$-meson–nucleus bound state energies and absorption widths
for various nuclei. The $\phi$-meson–nucleus potentials were calculated using a
local density approximation, with the inclusion of the $K\Kbar$ loop in the
$\phi$-meson self-energy. The nuclear density distributions, as well as the
in-medium $K$ and $\Kbar$ meson masses, were consistently calculated by
employing the quark-meson coupling model. Using the $\phi$-meson--nucleus
complex potentials found, we have solved the Klein-Gordon equation in momentum
space, and obtained $\phi$-meson--nucleus bound state energies and absorption
widths. Furthermore, we have studied the sensitivity of our results to the
cutoff parameter $\Lambda_{K}$ in the form factor at the $\phi K\Kbar$ vertex
appearing in the $\phi$-meson self-energy. We expect that the $\phi$-meson
should form bound states for all four nuclei selected, provided that the
$\phi$-meson is produced in (nearly) recoilless kinematics. This feature is
even more obvious in the (artificial) case where the absorptive part of the
potential is ignored. Given the similarity of the binding energies and widths
reported here, the signal for the formation of the $\phi$-meson--nucleus bound
states may be difficult to identify experimentally. Therefore, the feasibility
of observation of the $\phi$-meson–nucleus bound states needs further
investigation, including explicit reaction cross section estimates.

\section{Acknowledgements}

This work was partially supported by Conselho Nacional de Desenvolvimento
Cient\'{\i}fico e Tecnol\'ogico-CNPq, Grants No.~152348/2016-6 (J.J.C-M.),
No.~400826/2014-3 and No.~308088/2015-8 (K.T.), No.~305894/2009-9 (G.K.), and
No.~313800/2014-6 (A.W.T.), and Funda{\c c}\~{a}o de Amparo \`{a} Pesquisa do
Estado de S\~ao Paulo-FAPESP, Grants No.~2015/17234-0 (K.T.) and
No.~2013/01907-0 (G.K.). This research was also supported by the University of
Adelaide and by the Australian Research Council through the ARC Centre of
Excellence for Particle Physics at the Terascale (CE110001104), and through
Grant No.~DP151103101 (A.W.T.). J.J.C-M. also acknowledges the support given
by Red-FAE-CONACyT (M\'exico) to attend the workshop where this work was
presented.

\end{document}